\documentclass[prl,a4paper,superscriptaddress,twocolumn,showpacs,amsmath,amssymb,floatfix,amstex]{revtex4}

\usepackage{psfrag}
\usepackage[dvips]{graphicx}
\usepackage{bm,pstricks,longtable}

\input{epsf}

\begin{document}

\title{Quantum compacton vacuum}
\author{O.V.Zhirov}
\affiliation{\mbox{Budker Institute of Nuclear Physics, 
Novosibirsk 630090, Russia}}
\author{A.S.Pikovsky}
\affiliation{\mbox{Department of Physics and Astronomy, Potsdam University,
  Karl-Liebknecht-Str 24, D-14476, Potsdam-Golm, Germany}}
\author{D.L.Shepelyansky}
%\homepage[]{http://www.quantware.ups-tlse.fr}
\affiliation{\mbox{Laboratoire de Physique Th\'eorique (IRSAMC), 
Universit\'e de Toulouse, UPS, F-31062 Toulouse, France}}
\affiliation{\mbox{LPT (IRSAMC), CNRS, F-31062 Toulouse, France}}

%\date{\today}
\date{May 5, 2010}

\begin{abstract}
We study the properties of classical and quantum compacton chains
by means of extensive numerical simulations. 
Such chains are strongly nonlinear and their  classical dynamics 
remains chaotic at arbitrarily low energies. We show that the collective
excitations of classical chains are described by sound waves
which decay rate scales algebraically with the wave number
with a generic exponent value. The properties of the quantum chains
are studied by the quantum Monte Carlo method and it is
found that the low energy excitations are well described by effective phonon
modes with the sound velocity dependent on an effective Planck constant.
Our results show that at low energies the quantum effects 
lead to a suppression of chaos
and drive the system to a quasi-integrable regime of effective phonon modes.

\end{abstract}

%05.45.-a Nonlinear dynamics and chaos 
%63.20.Ry Anharmonic lattice modes 
%45.50.Jf Few and many-body systems

\maketitle

\section{I. Introduction}
The investigation of nonlinear chains, started by Fermi, Pasta and Ulam in
1955 \cite{fpu}, still remains an active and interesting area of research
which attracts a significant interest of nonlinear community
(see e.g. reviews \cite{flach,lepri,gallavotti,ruffo} and Refs. therein).
Usually, in such chains the nonlinear terms are relatively weak
compared to the linear ones and strong nonlinear effects appear only at
sufficiently high energy excitations.

However, there are also other types of chains where
the linear modes are absent and the dynamics is strongly nonlinear at
arbitrarily small energies. In such chains a time can be rescaled with energy
and hence the system always remains in a strongly nonlinear regime.
A prominent example is the Hertz lattice which describes
elastically interacting hard balls where the elasticity parameter
scales as a square-root of the displacement
\cite{nesterenko1983,fauve,chatterjee,porter} corresponding to
the nonlinearity index $n=5/2$.
Nesterenko \cite{nesterenko1983,nesterenko1985,nesterenko1994,nesterenko2001}
described a compact traveling-wave solution in the Hertz lattice
now known as compacton. A rigorous mathematical description of compactons
was given by Rosenau and Hyman \cite{rosenau1,rosenau2} for a class of
nonlinear partial differential equations (PDEs) with nonlinear dispersion.
A detailed analysis of compacton dynamics on lattices with various
nonlinearity index has been performed recently in \cite{ahnert}.
It was shown that on a finite lattice interactions between compactons lead to
chaotic dynamics characterized by a spectrum of positive Lyapunov exponents. 
A well known example of such compacton chain is the toy ``Newton's cradle'' 
\cite{ahnert}.
 
In the compacton lattice the dynamics is chaotic at arbitrarily small
energies when more than one compacton are exited. Therefore it is interesting to 
understand what happens in such quantum lattices
and what are the properties of quantum compacton vacuum. The studies of
such quantum chains are reported in this work. We use the quantum Monte Carlo
(QMC) method and the approach developed for the quantum Frenkel-Kontorova
model as it is described in \cite{qfk}. The aim of this work is to understand
the properties of quantum compacton vacuum and low energy excitations.
It is interesting to note that the recent progress with cold atoms allowed to
realize a quantum Newton's cradle \cite{dweiss} and to study
energy redistribution between atoms. The experimental progress
stimulated also theoretical studies of integrability and non-integrability in
one-dimensional atomic lattices at high energy excitations \cite{olshanii}. 
In contrast to that we study the properties of vacuum and low energy excitations
in quantum compacton lattices when the linear terms are absent and
nonlinearity is always strong in the classical case. Thus the classical
dynamics of such compacton chain is always chaotic \cite{ahnert}
(except a special case of one compacton moving in a lattice).
What are the properties of the quantum compacton chain is the subject of 
studies of this work.

The paper is organized as follows: 
the model description is given in Section II,
the properties of sound waves 
in the classical compacton chains are analyzed in Section III,
simple analytical estimates for the  compacton chain are
presented in Section IV, numerical results of the QMC are presented in Section
V and the results are summarized in Section VI.

\section{II. Model description}
The quantum compacton chain is described by the Hamiltonian
\begin{equation}
  \hat{H} = \sum_{l=1}^{M} \frac{1}{2} \hat{p}_l^2 + \frac{\alpha}{n} (
  \hat{x}_l - \hat{x}_{l - 1})^n
\label{eq1}
\end{equation}
where index $l$ marks $M$ particles in the chain and $n$ is 
the nonlinearity index.
Here $x_l$ gives the particle coordinate counted from the equilibrium distance
between particles which is taken to be $a$. For the Newton cradle $a$ is
given by the ball diameter. We use the dimensionless units in which the
particle mass is equal to unity and the momentum $p_l$ gives the particle
velocity. For the quantum problem the operators of momentum and coordinate 
have the usual commutator $[\hat{p}_l,  \hat{x}_{l'}] = -i \hbar
\delta_{l,l'}$ with a dimensionless Planck constant $\hbar$. We assume the
periodic boundary conditions with $x_l=x_{l+M}$, $p_l=p_{l+M}$
or fixed ends boundary conditions. The later is a specific case of
even number of particles $M=2N$. We note that the particles are
distinguishable since they are located at well defined positions.

Let us remind few known results for the harmonic chain at $n=2$.
By a canonical transformation to normal modes
$   \hat{P}_k = \frac{1}{\sqrt{N}} \sum_l \hat{p}_l \mathrm{e}^{- 2 \pi \mathrm{i} l k/M}$, 
$\hat{Q}_k = \frac{1}{\sqrt{N}} \sum_l \hat{x}_l \mathrm{e}^{2 \pi \mathrm{i} l k / M}$
the Hamiltonian (\ref{eq1}) takes the form
\begin{eqnarray}
 \hat{H} & = & \frac{1}{2} \sum_k 
\left( \left| \hat{P}_k \right|^2 + {\omega_k}^2 \left| \hat{Q}_k \right|^2 \right)
\label{eq2}
\end{eqnarray}
 with the normal mode frequencies
$  \omega_k = 2 \bar{\omega} \sin (q_k/2)$, $ \bar{\omega}=\sqrt{\alpha}$ and wave numbers
$q_k = \pi  k/N$, $ k = 1, \ldots, N - 1$.
Here and below we use the fixed boundary conditions with $x_{l=0}=x_{l=N}=0$
and $N$ particles. 

It is convenient to introduce sine modes via relations
$ \hat{S}_k  =  \sqrt[]{\frac{2}{N}} \sum_l \sin (q_k l) \hat{x}_l$,
$ \hat{x}_l  =  \sqrt[]{\frac{2}{N}} \sum_k \sin (q_k l) \hat{S}_k$.
The vacuum state of the chain is a product of vacuum states of all  modes.
For any mode in the vacuum state one has an average of mode energy $\hat{U}_k$ being
$\left\langle \hat{U}_k \right\rangle =
\omega^2_k \left\langle \hat{S}^2_k \right\rangle / 2 = \hbar \omega_k / 4$ and hence
\begin{equation}
  \left\langle \hat{S}_k \hat{S}_{k'} \right\rangle = \frac{\hbar}{2 \omega_k}
  \delta_{k k'} . 
\label{eq3}
\end{equation}
Thus different harmonics are independent and as a result the squared deviation 
from equilibrium for a particle $l$ is 
$ \left\langle \hat{x}_l^2 \right\rangle = \frac{2}{N} \sum_{k = 1}^{N - 1}
  \sin^2 (q_k l) \left\langle \hat{S}^2_k \right\rangle = 
\frac{\hbar}{N} \sum_{k = 1}^{N - 1} \sin^2 (q_k l) / \omega_k$.
For the central particle $l=N/2$ the displacement diverges logarithmically with the chain
length 
$\left\langle \hat{x}_l^2 \right\rangle = 
\frac{\hbar}{4 n} \sum_{k = 1}^{N -1} 1 / \sin (q_k / 2) \approx
 \frac{\hbar}{\pi} \ln (2 / q_{\min})$, where $q_{\min} \equiv q_{k = 1} = \pi / N$.
For $\hbar=1$ the displacement is of the order of unity for
$N \sim 30$. We will assume that such displacements are small compared to the
distance $a$ between particles.

The spacial correlator of the chain can be also explicitly calculated
as $\left\langle \hat{x}_l \hat{x}_{l + \Delta} \right\rangle  =  \frac{2}{N} \sum_{k =
  1}^{N - 1} \sin (q_k l) \sin (q_k (l + \Delta)) \left\langle \hat{S}_k^2 \right\rangle $
where the brackets  note the quantum average. Using (\ref{eq3}) we obtain
after summation over all $l$
\begin{equation}
  \left\langle \hat{x}_l \hat{x}_{l + \Delta} \right\rangle_l \approx \frac{\hbar}{N}
  \sum_{k = 1}^{N - 1} \frac{\cos (q_k \Delta)}{4 \sin (q_k / 2)}
  \label{eq4}
\end{equation}

At finite temperature $T$ the relation (\ref{eq3}) is modified for the usual 
expression for bosons
\begin{equation}
  \left\langle \hat{S}_k^2 \right\rangle = \frac{\hbar}{\omega_k} \left( \frac{1}{2}
  + \frac{1}{\exp (\hbar \omega_k / T) - 1} \right) .
\label{eq5}
\end{equation}
With this form of $\left\langle \hat{S}_k^2 \right\rangle$ one can obtain the
expression for the correlator $\left\langle \hat{x}_l \hat{x}_{l + \Delta} \right\rangle$
at finite temperature.

The properties of the chain can be also characterized by the static form factor
defined as 
\begin{equation}
  F (q) = \left\langle \left| \sum_l \exp \left( \mathrm{i} (a \cdot l + \hat{x}_l) q /
  \hbar \right) \right|^2 \right\rangle \; ,
\label{eq6}
\end{equation}
where $q_{}$ can be viewed as a momentum transfer during a process of photon
scattering, and $a$ is the spacing of the chain lattice.

Before to start the studies of the quantum compacton problem at $n>2$
we consider in the next Section
the propagation of sound waves in the classical chain at $n=4$.

\section{III. Sound waves in classical compacton lattices}

To find the spectrum of sound waves and their decay rates we 
simulate numerically the dynamics of classical system (\ref{eq1}) at $n=4$, $\alpha=1$ with
the periodic boundary conditions.
The simulations are done with a   Runge-Kutta-Nystrom method  with
time  step $0.02$. Initially a random distribution of momenta $p_l$ is seeded, so
that the total energy per particle is 1. 
After a transient of time $T_{trans}=10$ a perturbation of the type
$x_l \to x_l+\varepsilon \cos q_k l$ is imposed with $\varepsilon=0.1$. 
The wave number $q_k$ is changed in the range $0<q_k<\pi$. 
The corresponding Fourier mode of $x_l$ changes in time  approximately as
 $f_q \propto  e^{-\gamma_q t}\cos\omega_q t$. This time dependence is
obtained from an ensemble of particles.  We use up to $10^7$ particles
in an ensemble to obtain good averaging of statistical fluctuations.

\begin{figure}[htbp]
\begin{center} 
\includegraphics[width=.95\linewidth]{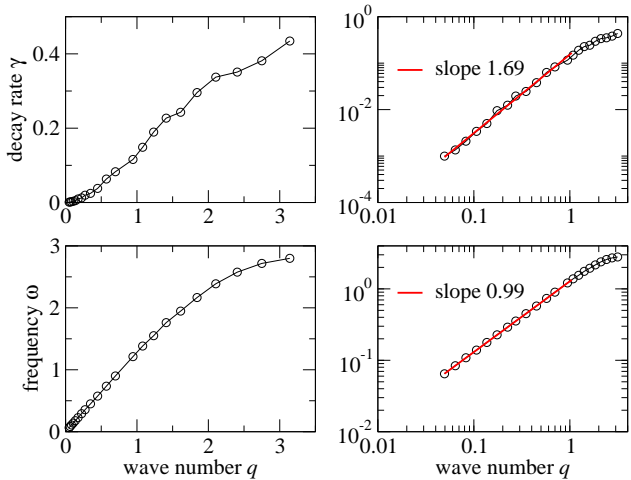}
\end{center} 
 \vglue -0.3cm
\caption{(Color online) Bottom panels: 
spectrum of waves $\omega(q)$ in the classical compacton lattice (\ref{eq1})
at $n=4$, $\alpha=1$ and energy per particle $\langle E_l \rangle$ equal to unity, 
at small wave numbers $q$ the spectrum is close to a linear law of sound waves
$\omega \propto q$. Top panels: decay rate of sound waves $\gamma(q)$,
at small $q$ we have $\gamma \propto q^\beta$ with $\beta \approx 1.69$.
Left panels are in normal scale, right panels are in log-log scale,
fits are shown by straight lines with indicated slopes.
}
\label{fig1}
\end{figure}

The numerical results are shown in Fig.~\ref{fig1} for the
spectrum of sound waves $\omega(q)$ and their decay rates $\gamma(q)$.
The frequency spectrum is close to the spectrum of sound in
a harmonic lattice  $\omega(q)=2 \bar{\omega} \sin q/2$ with
$\bar{\omega} \approx 1.4$. At small $q$ we have the spectrum of sound waves.
The effective velocity of sound 
for phonon like excitations is given by
$\bar{\omega} \approx 1.4 \langle \alpha E_l\rangle ^{1/4}$.
Here the dependence on  a given average
energy per particle $\langle E_l \rangle$
appears since
a typical frequency of particle oscillations is proportional to $\langle E_l \rangle^{1/4}$. 
The speed of sound 
in the physical lattice with distance $a$ between particles
is $c=\bar{\omega} a$. 
The decay rate of these waves drops algebraically
with the decrease of the wave vector $q$ as $\gamma \propto q^{\beta}$.
We obtain the value $\beta \approx 1.69$ that is close to the generic exponent
$\beta =5/3$ for the decay rate in nonlinear lattices
(see e.g. \cite{lepri}). Thus even if the dynamics inside a compacton lattice
is strongly chaotic (see \cite{ahnert}) the long wave oscillation properties of the
whole lattice are well described by effective sound waves.
In a certain sense the situation is similar to sound in a gas media: each
particle moves chaotically but the collective long wave excitations are well
described by sound waves. 

It is interesting to note that recently a localization of 
sound waves in a random three-dimensional elastic network 
of metallic balls has been observed experimentally in \cite{skipetrov}.
Such a system can be viewed as a random three-dimensional Newton's cradle.
However, our studies here are restricted to the one-dimensional case.

\section{IV. Simple estimates for quantum compacton chains}

The sound velocity $c$ in a gas is given by the derivative of pressure $p$ over 
the gas mass density $\rho$ at fixed entropy $S$ (adiabatic process):
$ c^2 = \left( \frac{\partial p}{\partial \rho} \right)_S$. 
Since the pressure is proportional to the force
$p \propto \partial U/\partial x_l$, hence,
$c^2 \propto\partial^2 U / \partial^2 x_l$. This leads to a simple estimate
for the sound velocity based on the virial theorem according to which
 $2 \left\langle K \right\rangle = n \left\langle U \right\rangle$,
where $K$ and $U$ are particle kinetic and potential energies and brackets
mark their average values. Since the temperature is proportional to the
kinetic energy we have
$T \sim \left\langle K \right\rangle = 
n \left\langle U \right\rangle = \alpha \left\langle (x_i - x_j)^n \right\rangle
\sim \alpha (\Delta x)^n$ where $\Delta x$ is an average displacement of a
particle. This gives 
\begin{equation}
c^2 = \omega^2 a^2 \sim a^2 U^{''}(\Delta x) 
\sim a^2 \alpha^{2/n} T^{1-2/n} \; .
\label{eq7}
\end{equation}
For the classical case this expression agrees with the above 
analytical results for $\bar{\omega}$  
for $n=2,4$. 
 
For the quantum compacton chain we can use the Heisenberg uncertainty relation
$p \sim \hbar / \Delta x$ for the minimization of the ground state energy 
$E = \hbar^2 / 2
(\Delta x)^2 + \alpha (\Delta x)^n / n$ that gives $\Delta x \sim (\hbar^2 /
\alpha)^{1 / (n + 2)}$, and $\omega^2 \sim \alpha (\Delta x)^{n - 2} =
\alpha^{4 / (n - 2)} \hbar^{2 (n - 2) / (n + 2)}$. As a result the sound
velocity of the quantum chain is
\begin{equation}
  c^2 \sim a^2 \alpha^{4 / (n - 2)} \hbar^{2 (n - 2) / (n + 2)}
\label{eq8}
\end{equation}

For $n=2$ this velocity is independent of $\hbar$ but for
$n>2$ it decreases with $\hbar$ that corresponds to the decrease
of the ground state energy of a nonlinear oscillator.
The properties of the ground state are analyzed in the numerical
simulations presented in the next Section.

\section{V. Numerical results of quantum Monte Carlo}

\subsection{A. Method description}

For our numerical simulations of quantum chain (\ref{eq1})
we use the Metropolis algorithm (MA) \cite{metropolis} in the Euclidean time
$\tau$ related to the system temperature $T=\hbar/\tau$. 
The simulations are done in the same way as for the studies of the quantum
Frenkel-Kontorova model described in \cite{qfk}. A general description
of this QMC method can be found in \cite{girvin}.

The paths in the discretized Euclidean time are generated by the statistical sum
\begin{eqnarray}
\label{eq9}
  \sum_{\{x_{l,j} \}} \exp \left\{ - \sum_{l, j} \left[ \frac{1}{2 \Delta
  \tau} (x_{l, j} - x_{l, j - 1})^2 
 + \frac{\Delta \tau}{2 n} (x_{l, j} - x_{l - 1, j})^n \right] \right\}
\end{eqnarray}
which links the problem to a  statistical mechanics for
configuration distribution of some lattice of $N \times N_{\tau}$ size
where $N$ is number of particles and $N_{\tau}=\tau/\Delta \tau$
is number of discrete steps of size $\Delta \tau$  in the  Euclidean time
interval $\tau$. As usual the periodic boundary conditions are used
in this time with $x_l(\tau_j+\tau) = x_l(\tau_j)$ and $\tau_j=j \Delta \tau$.
In our numerical studies we  use up to $N_\tau=1000$, $\tau=200$ and up to
$2 \cdot 10^6$ Metropolis updates. We fix $\alpha=1$ for numerical
simulations.

The numerical simulations assume a generation of configurations $\{x_{k, j} \}$ with
probability proportional to their weights in the statistical sum. 
The Metropolis method looks very efficient,
providing an update step gives 
rather large modifications for a give site. However,
corresponding modifications are local and are dominated by the nearest
neighbor sites, that results in a significant slowdown for
long wave configurations. Thus it is useful to combine
the Metropolis method with the microcanonic dynamics (MCD) method.
The MCD method is a noiseless algorithm and it works as follows:
 all variables
$x_{l, m}$ are  considered as some coordinates, and the sum
\begin{equation}
\label{eq10}
  \sum_{l,j} \left[ \frac{1}{2 \Delta \tau} (x_{l, j} - x_{l, j - 1})^2 +
  \frac{\Delta \tau}{2 n} (x_{l, j} - x_{l - 1, j})^n \right] \equiv
  \mathcal{U}(x)
\end{equation}
as a potential energy of a certain system. Then, a set of auxiliary
momentum variables $\{{\cal P}_{l,j} \}$ is added and the equations of motion are
solved numerically in an auxiliary update ``time'' variable $u$:
\begin{eqnarray}
\label{eq11}
  \partial{{\cal P}}_{l, j}/\partial{u} & = & - \frac{\partial \mathcal{U}}{\partial x_{l, j}} =
  \nabla_{l, j} \mathcal{U}(x) \nonumber\\
  \partial{x}_{l, j}/\partial{u} & = & {\cal P}_{l, j} \; .
\end{eqnarray}
With this dynamical description the system evolves 
over some iso-energy hypersurface in the phase
space, and we get an ensemble of configurations.

\begin{figure}[htbp]
\begin{center}
  \includegraphics[width=.95\linewidth]{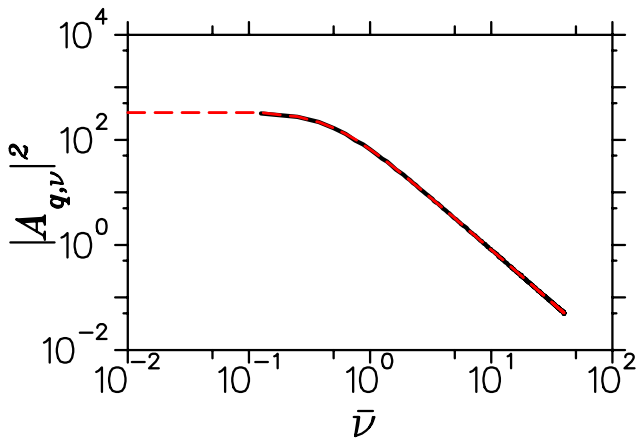}
\end{center}
\vglue -0.3cm
  \caption{(Color online) Typical data for the Fourier harmonics $|A_{q_j, \nu_m} |^2$ and its
  fit by the dependence (\ref{eq14}); black curve shows numerical data, dashed
  red/gray
  curve shows the fit (curves are overlapped). Data
  correspond to the linear chain at $n=2$, $\hbar = 1$, \ $q_j = 10\pi/N$, $N = 64$,
  $\tau_{\max} = 100$, $N_{\tau} = 1000$.}
\label{fig2}
\end{figure}

However, an obvious disadvantage of such a method is that one needs  to solve the
differential equations numerically with a step which 
becomes smaller for decreasing
$\Delta \tau$, due to the terms $\mathcal{U}(x)$ with $\Delta \tau$
in the denominator. But even worse, these terms act as some noise that
reduces the relaxation rate along $l$ (space dimension). Thus, in order to
accelerate the  relaxation processes we introduce into $\mathcal{U}(x)$ a parameter
$C_K$
\begin{equation}
\label{eq12}
  \mathcal{U}(x ; C_K) = \sum_{l,j} \left[ \frac{C_K}{2 \Delta \tau} (x_{l,j}
    - x_{l, j - 1})^2 + \frac{\Delta \tau}{2 n} (x_{l, j} - x_{l - 1, j})^n
  \right] .
\end{equation}
Then the update step   is organized as follows:\\
a) the fast mixing stage, with $C_K = \Delta \tau$, at which the
  smallness of denominator is canceled, here the MCD method is applied during the
  auxiliary ``time'' interval $u \sim 100$;\\
b) return to  $C_K = 1$ and application of the Metropolis updates
at maximum 400 updates.

Such a combined approach allows us to reduce significantly the number of Metropolis
updates and to have more rapid numerical simulations. We checked that
both methods (only MA steps and MA steps combined
with the MCD method) give the same results.

\subsection{B. Quantum excitations above the compacton vacuum}

The ensemble of quantum paths, obtained by the numerical methods described
above, determines the properties of the vacuum ground state of the system. 
Averages of various quantities over this ensemble give
the corresponding average observables at this state. 
However, the study of fluctuations of quantum
paths allows us to extract a more interesting information
about  the spectrum of low energy elementary excitations.
\begin{figure}[htbp]
\begin{center}
  \includegraphics[width=.85\linewidth]{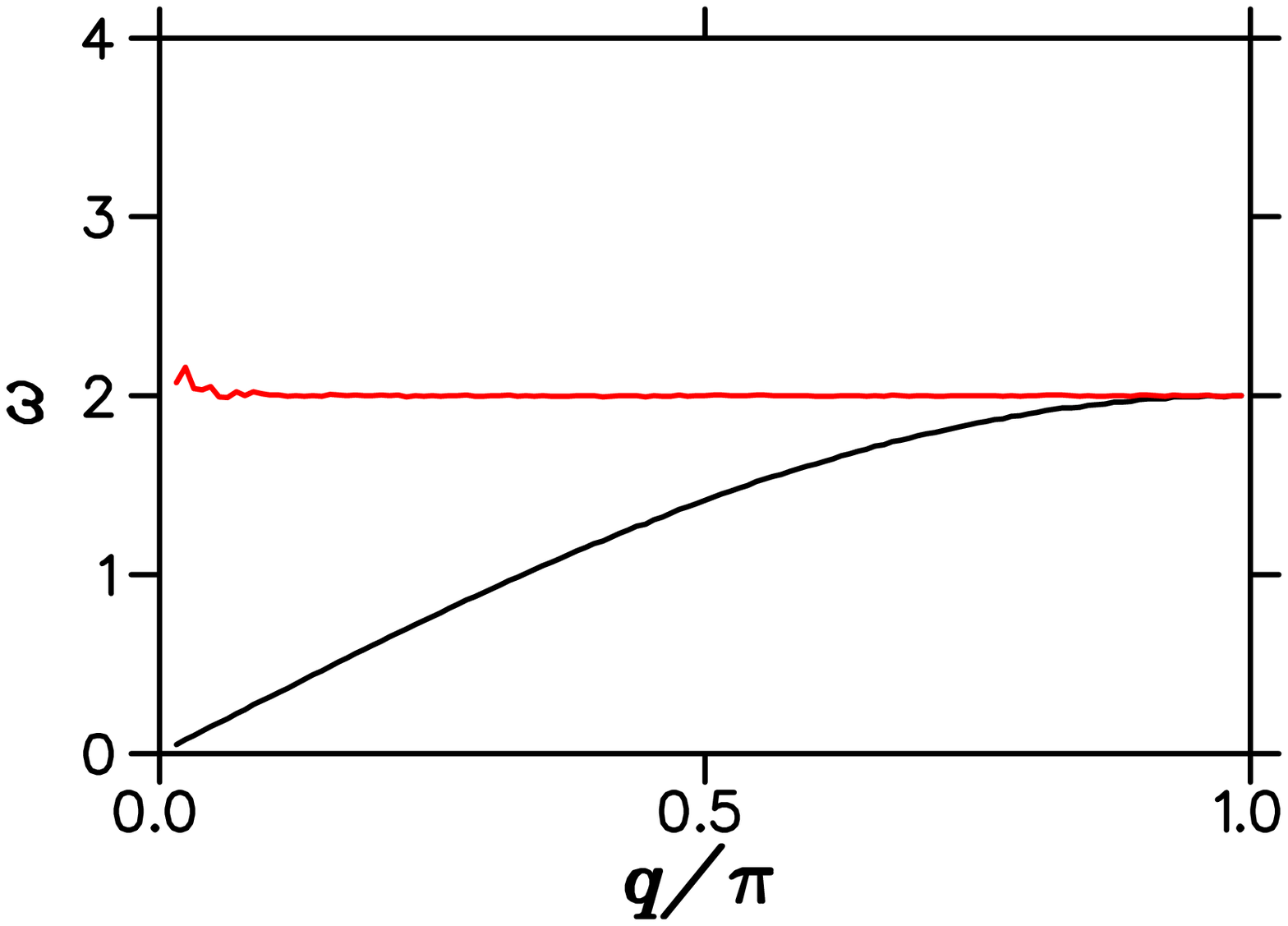}\\
  \includegraphics[width=.85\linewidth]{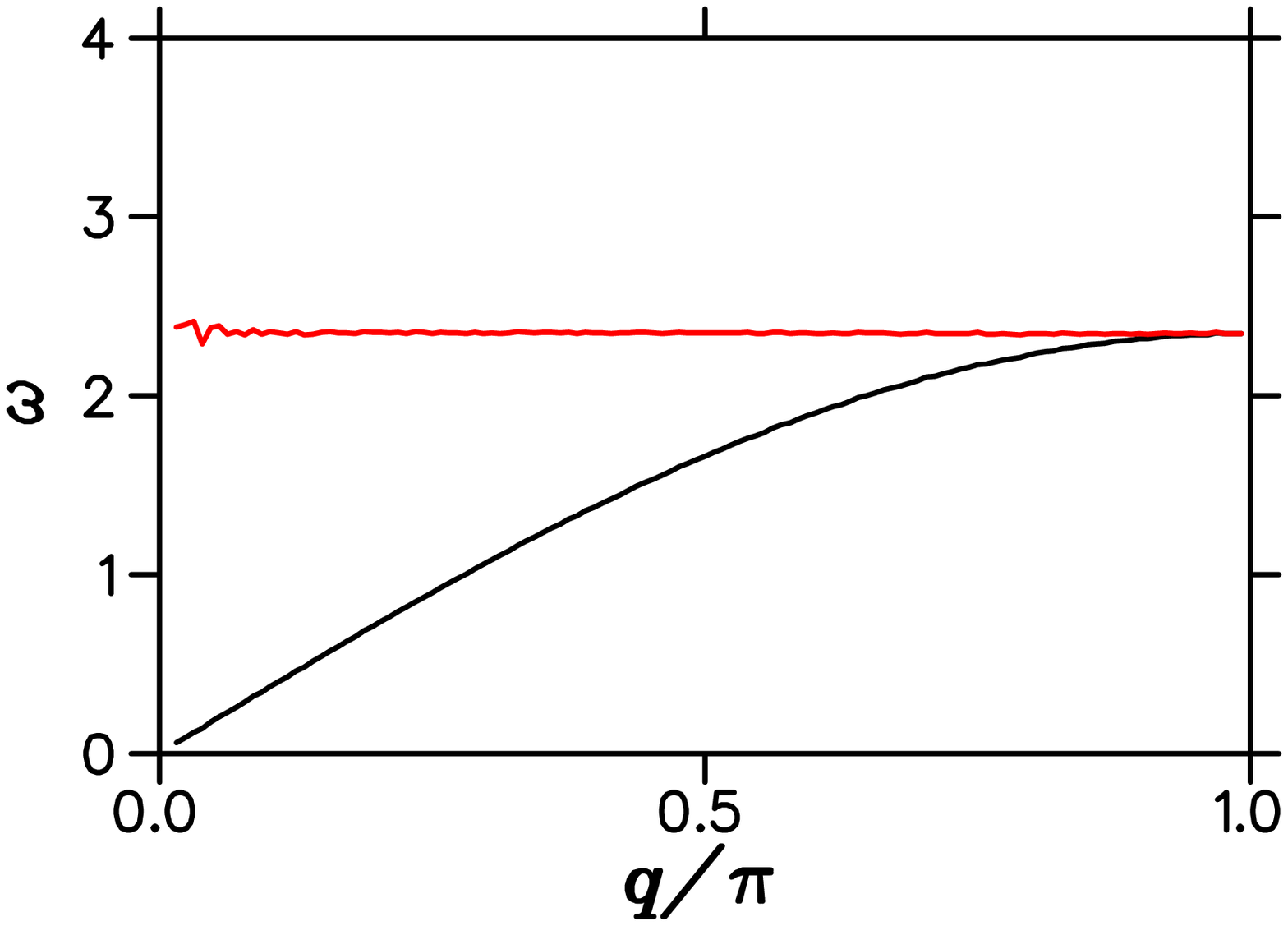}\\
  \includegraphics[width=.85\linewidth]{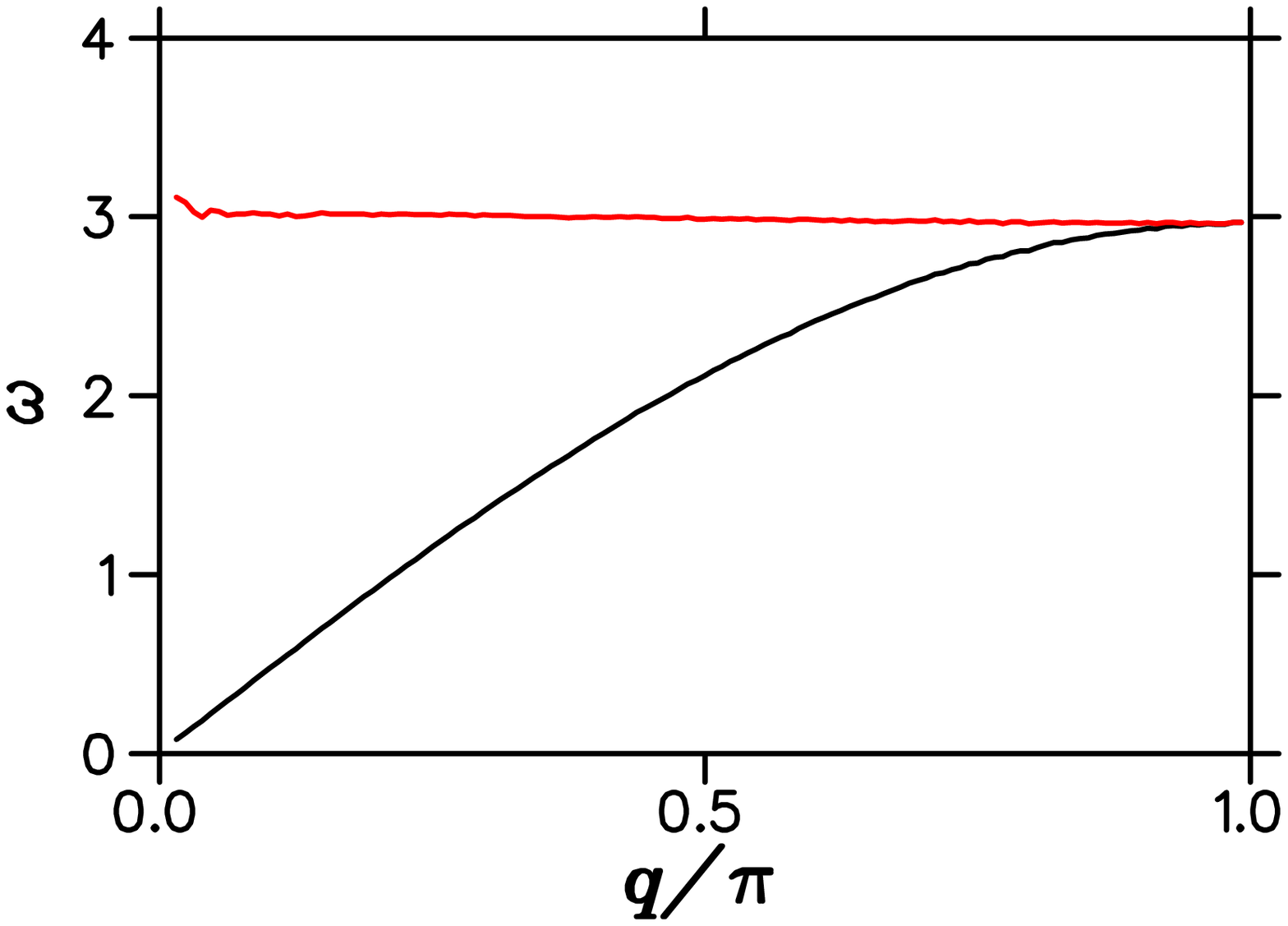}
\end{center}
\vglue -0.3cm
  \caption{(Color online) Spectra of quantum phonon modes shown by black bottom curve and
   obtained from QMC simulations for
  $n=2$ (top panel), $n=4$ (middle panel) and $n=8$ (bottom panel); top
   red/gray curve shows the ratio $\omega / \sin(q/2)$ for each panel. The data are
    obtained at $\hbar=1$, $N=128$, $N_{\tau} = 1000$, $\tau_{\max} = 200$,
    the simulations are done with MA using $2 \cdot 10^6$ updates.}
\label{fig3}
\end{figure}

To extract this information we consider the Fourier harmonics of quantum paths
\begin{equation}
\label{eq13}
  A_{q_j, \nu_m} = \sum_{l, k} x_{l, k} \sin (q_j l) \exp ({\mathrm i} \mathrm{ \nu_m}
  \tau_k)
\end{equation}
where $q_j = \pi j / N$, $j = 1, \ldots, N - 1$, $\tau_k=k \Delta \tau$,
$\nu_m = 2 \pi m / \tau_{\max}$, $m = 0, \ldots, N_{\tau} / 2$. 
One expects a Lorentzian distribution
in frequency $\omega$
for exponential decay of quasiparticle excitations in the imaginary time:
\begin{equation}
\label{eq14}
  \left\langle |A_{q_j, \nu_m} |^2 \right\rangle = \frac{\hbar}{2} 
  \frac{1}{\omega^2 + \bar{\nu}_m^2} .
\end{equation}
Here we use the renormalized frequency
$ \bar{\nu}_m = \frac{2}{\Delta \tau} \sin (\nu_m \Delta \tau / 2) $ with the
sine term appearing due to the discreetness of time steps. The fit of data
for $ A_{q_j, \nu_m}$ allows to find the spectral dependence $\omega = \omega (q_j)$
and thus to determine the dispersion law of elementary quantum excitations.
A typical example of such a fit is shown in Fig.~\ref{fig2}.

The spectrum of low energy excitations extracted via such a procedure is shown
in Fig.~\ref{fig3} for $n=2,4,8$. For $n=2$ the data reproduce 
the theoretical result for a harmonic chain with $\bar{\omega}=\omega(q)/(2\sin(q/2))=1$.
For $n=4,8$ this form of spectrum is preserved with a moderate
renormalization of $\bar{\omega} \approx 1.2, 1.5$ respectively.
This shows that even if the classical compacton chain is fully chaotic
the quantum compacton vacuum is rather regular and is characterized by
the phonon type excitations rather similar to the case of a harmonic chain.

\begin{figure}[htbp]
\begin{center}
  \includegraphics[width=.95\linewidth]{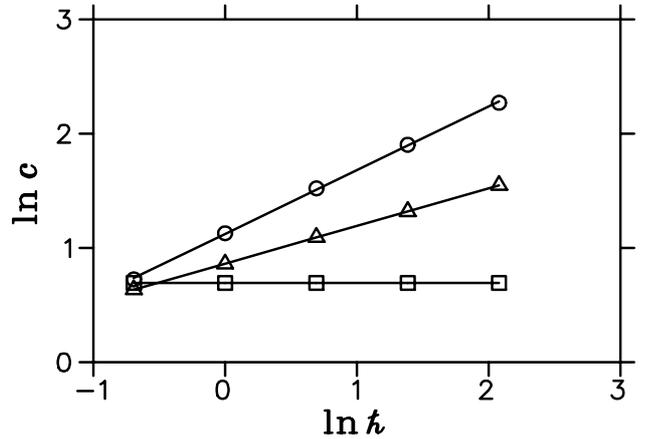}
\end{center}
\vglue -0.3cm
  \caption{The dependence of sound velocity $c$ \ on $\hbar$. Squares,
  triangles and circles correspond to $n = 2$, $4$ and $8$, respectively.
  The straight lines show the fit dependence (see text).
  Parameters of simulations are: $N = 64$, $\tau_{\max} = 100$, 
  $N_{\tau} = 1000$. Logarithms are natural.}
\label{fig4}
\end{figure}

\begin{figure}[htbp]
\begin{center}
  \includegraphics[width=.95\linewidth]{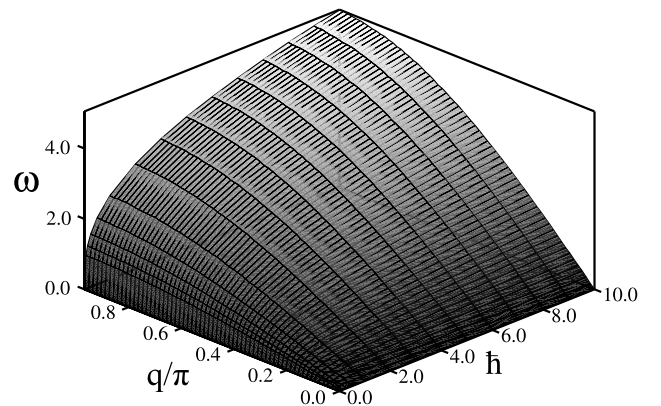}
\end{center}
\vglue -0.3cm
  \caption{Spectra of quantum phonon modes for the quartic chain at $n=4$. Data
  correspond to $N = 64$, $\tau = 100$, $N_{\tau} = 1000$, and to the interval
  $\hbar = 0.063 - 10$, \ extrapolated smoothly from $\hbar = 0.063$ to 
  $\hbar = 0$. }
\label{fig5}
\end{figure}

Our data show that $\bar{\omega}$ varies with $\hbar$ and $n$.
This leads to the dependence of the sound velocity $c=d\omega/d q$
on these two parameters.  The numerically obtained dependence of $c$ on
$\hbar$ is shown in Fig.~\ref{fig4}. The fit by an algebraic 
dependence $c \propto \hbar^\eta$ gives $\eta=0.33; 0.56$
for $n=4,8$ respectively. These numerical values are close to the
theoretical power from Eq.~(\ref{eq8}) with $\eta=(n-2)/(n+2)$
corresponding to $\eta=1/3, 3/5$ for these $n$ values.
The global variation of the whole spectrum $\omega(q)$
with $\hbar$ is shown in Fig.~\ref{fig5} for $n=4$.

\begin{figure}[htbp]
\begin{center}
  \includegraphics[width=.45\linewidth]{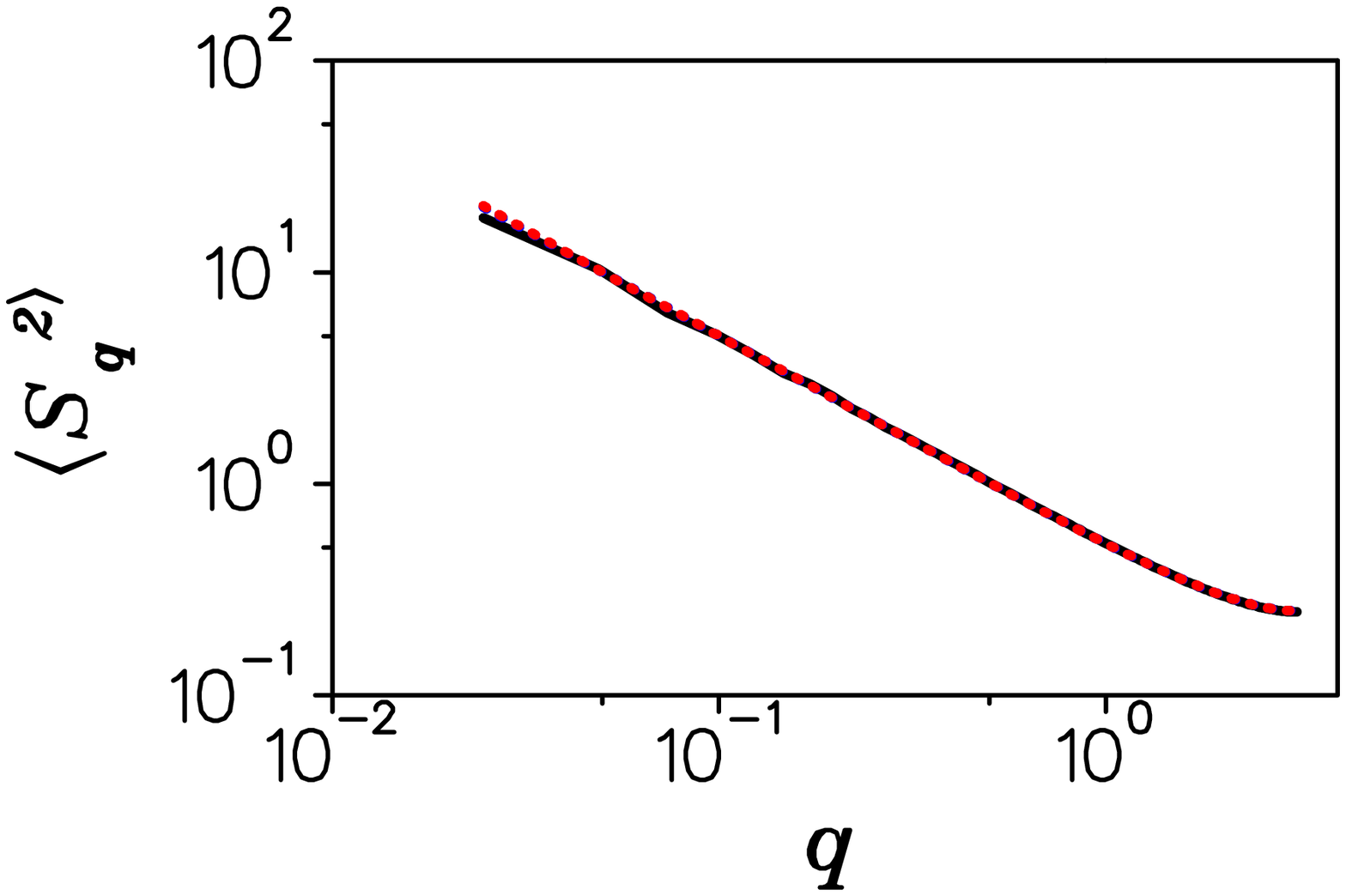}
  \includegraphics[width=.45\linewidth]{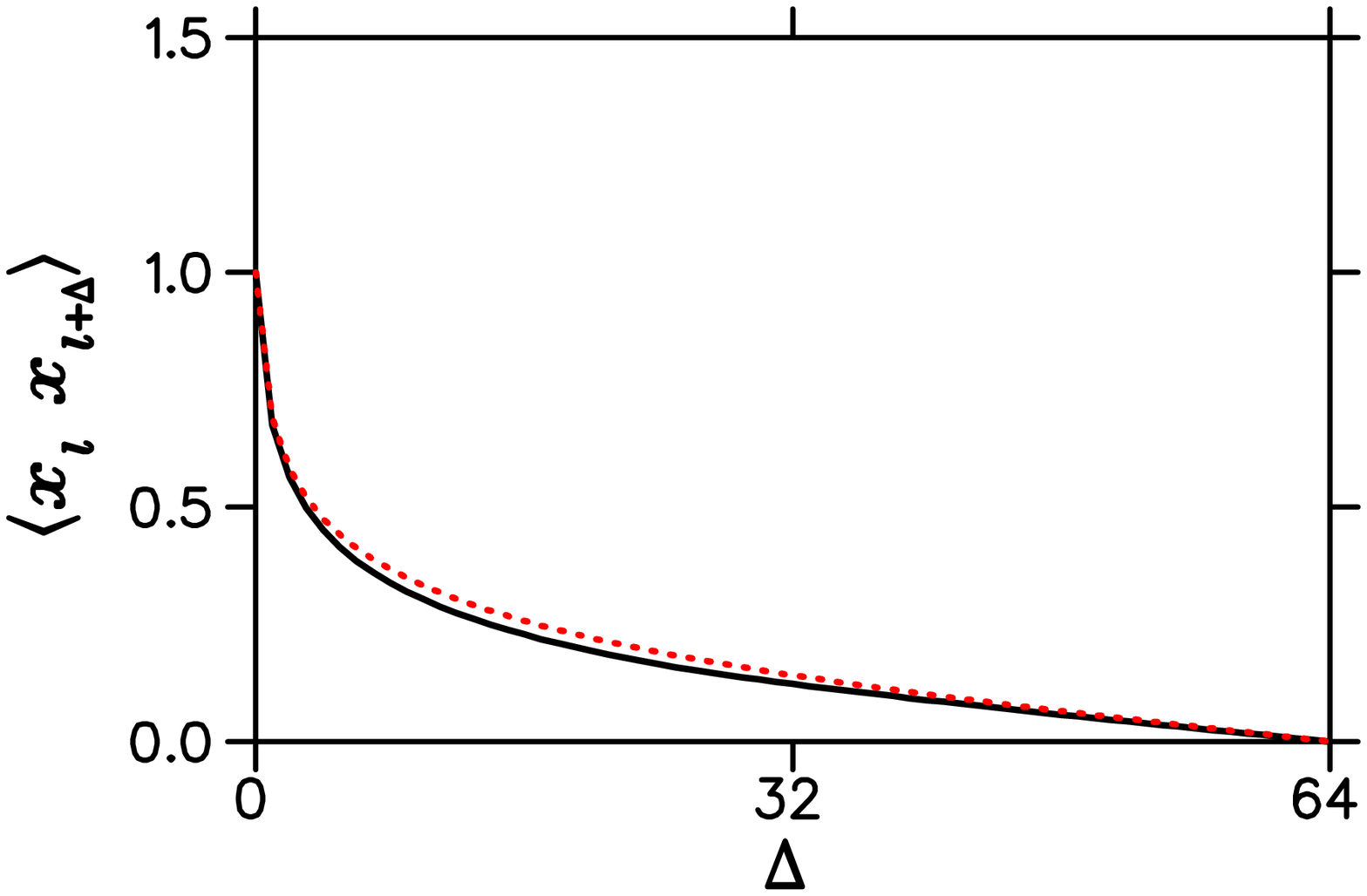}\\
  \includegraphics[width=.45\linewidth]{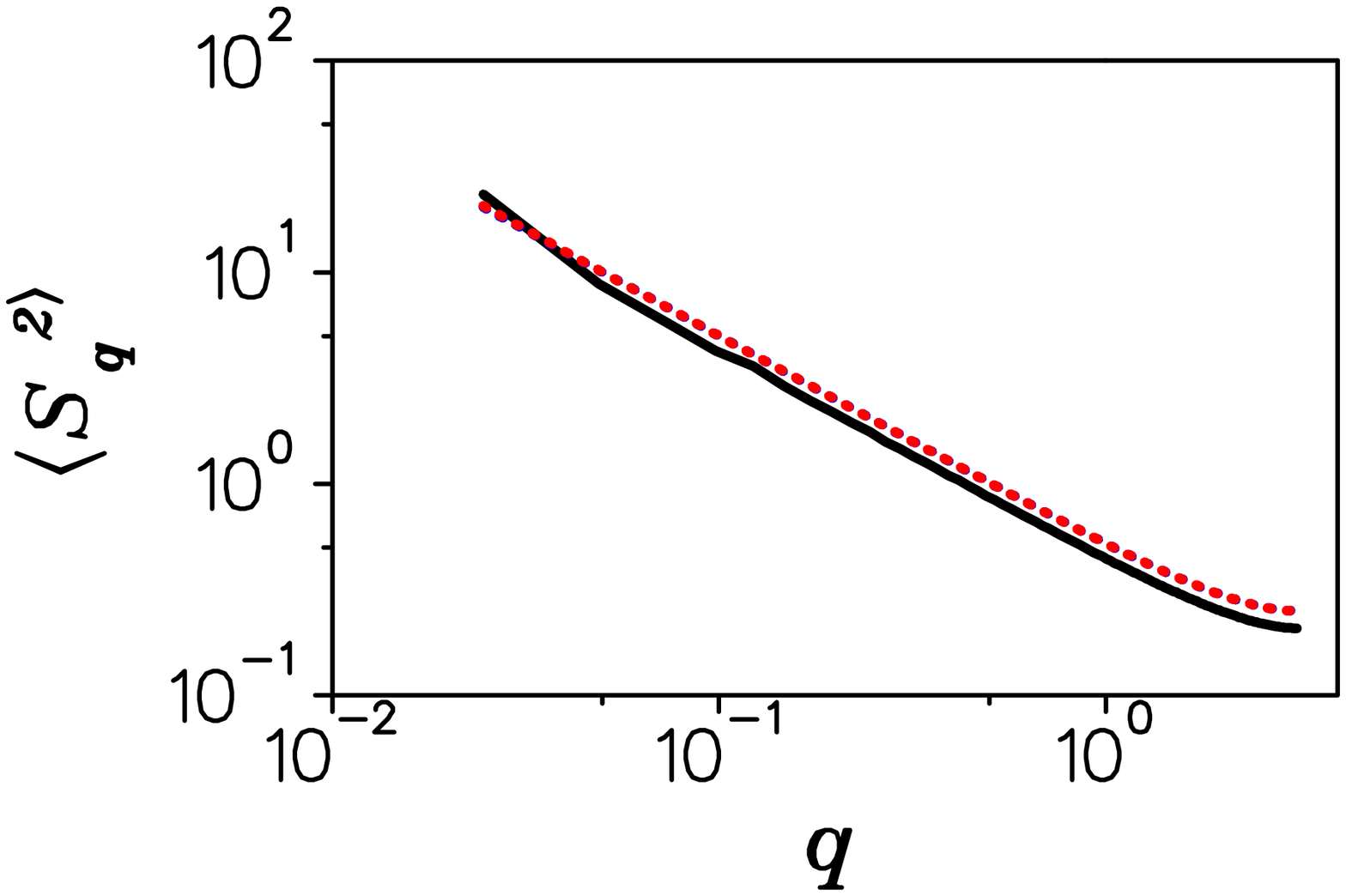}
  \includegraphics[width=.45\linewidth]{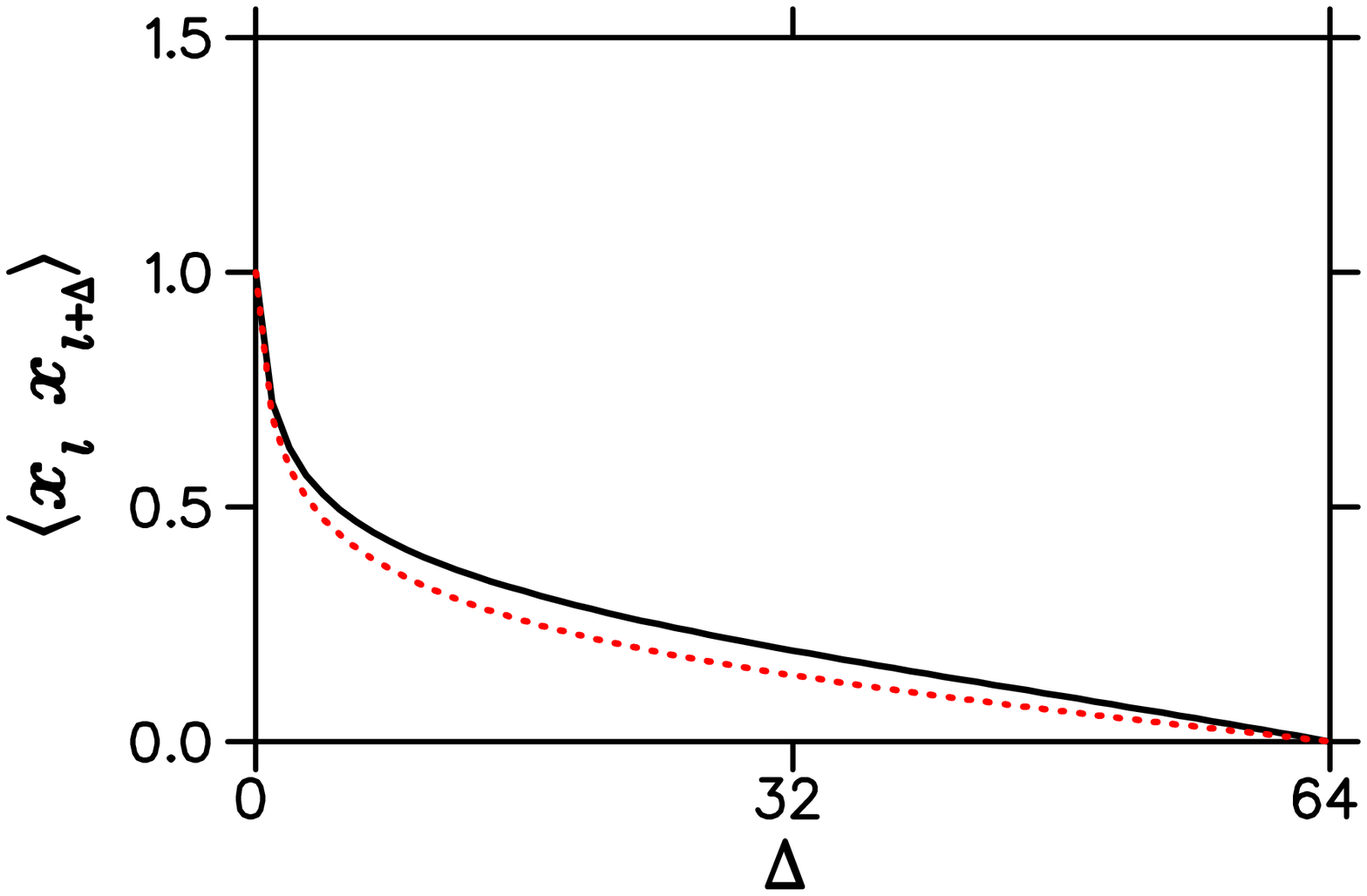}\\
  \includegraphics[width=.45\linewidth]{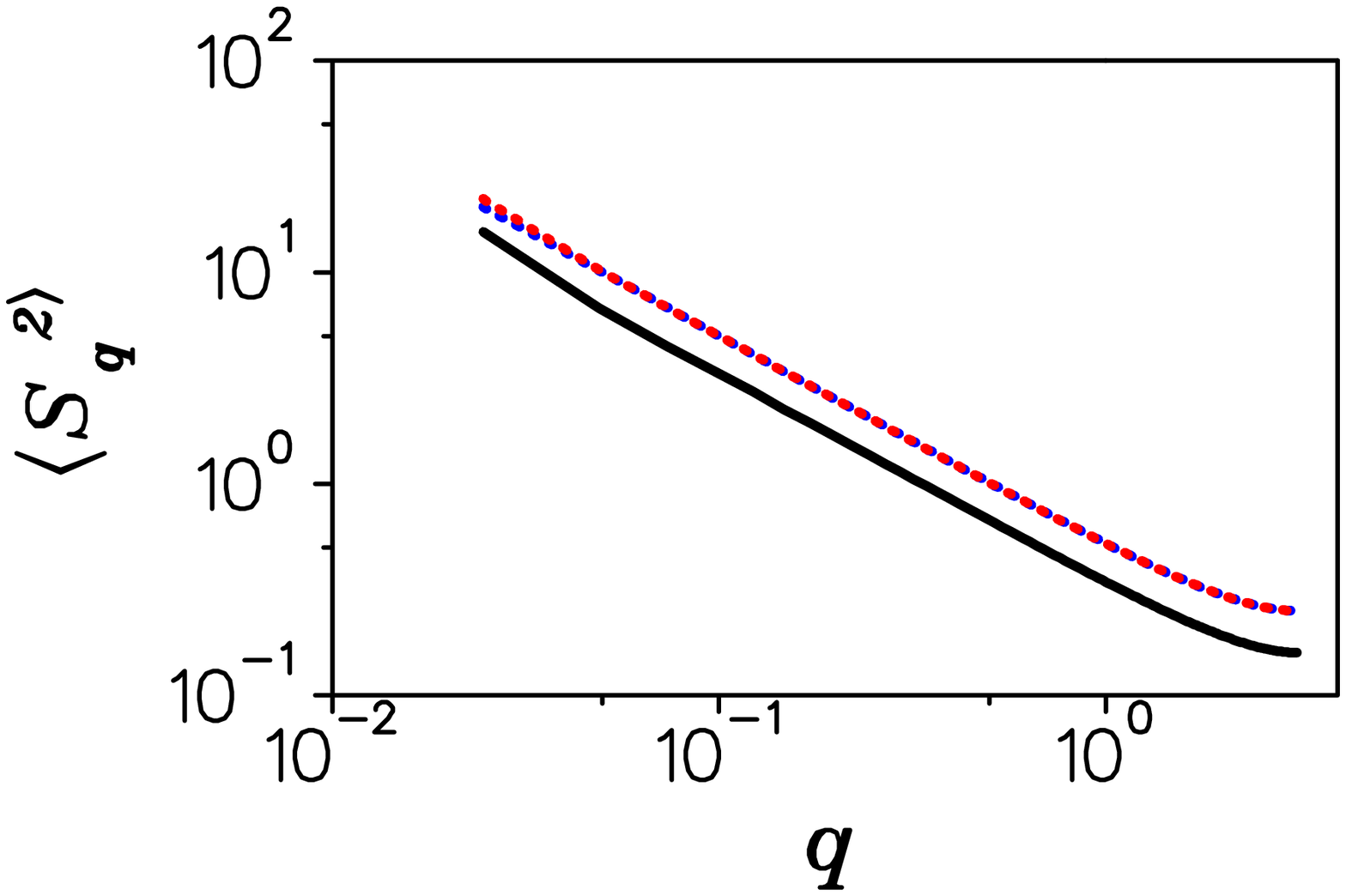}
  \includegraphics[width=.45\linewidth]{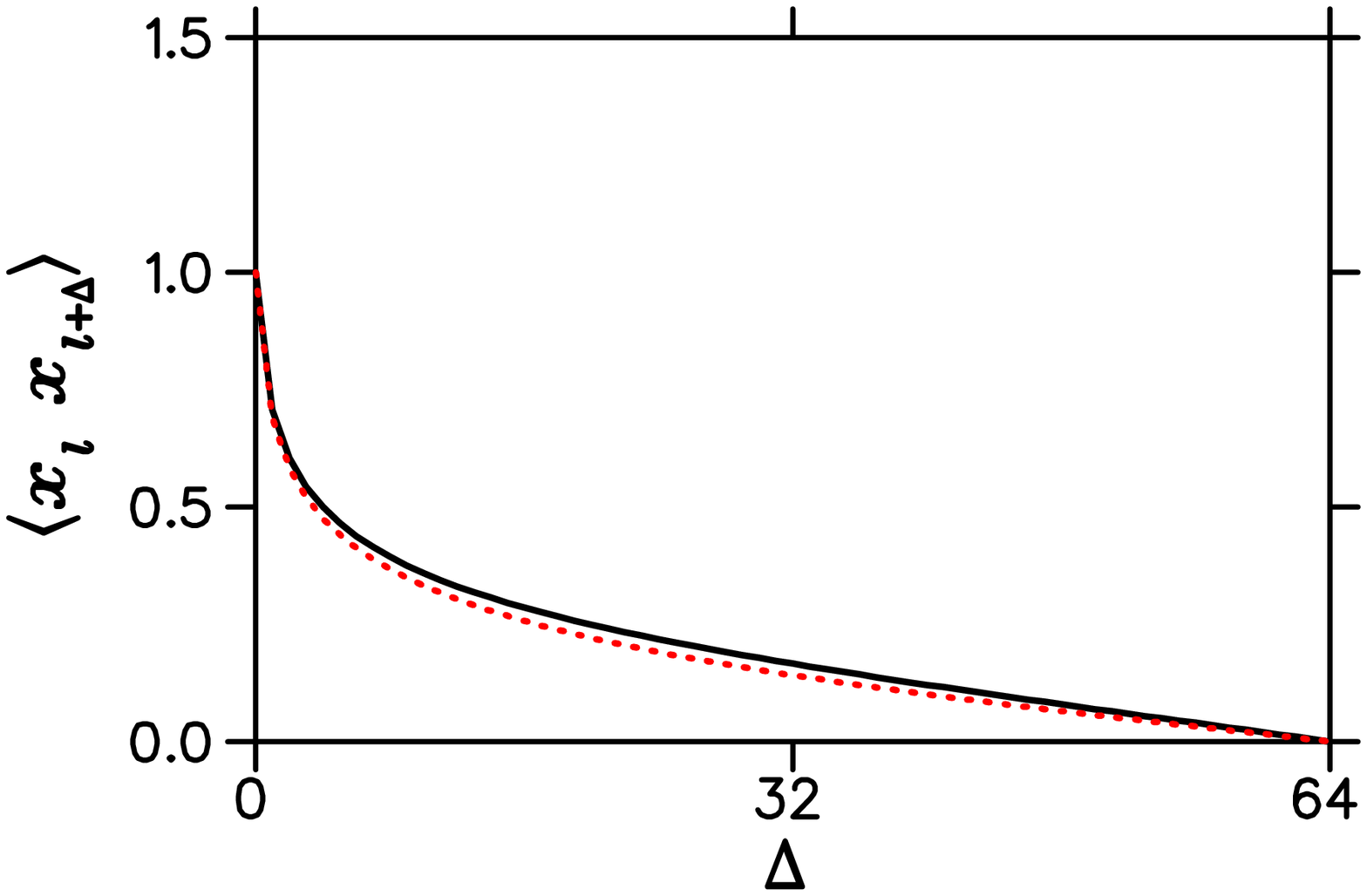}
\end{center}
\vglue -0.3cm
  \caption{(Color online) Left column: Amplitudes of phonon modes, see Eq.(\ref{eq3}), for
  $n = 2$ (top), $4$ (middle), $8$ (bottom). 
  Right column: Normalized correlation function, see Eq.(\ref{eq4}), 
  for $l = 64$ (central particle) for the same order of panels.
  Other parameters $\hbar = 1$, $N=128$, $N_{\tau} = 1000$, 
  $\tau = 200$. Simulations include $2 \cdot 10^6$ updates. Red/gray and blue/black dotted 
  curves
  give the theoretical expectations for harmonic chain with temperature 
 $T = \hbar / \tau$ and $T = 0$, respectively (curves overlap).}
\label{fig6}
\end{figure}

Trying to find deviations from a harmonic chain behavior we compute
numerically the amplitudes of phonon modes $ S_q^2$ (see Eq.~(\ref{eq3}) with $q=k/(N-1)$)
and the correlation function 
$ \left\langle x_l x_{l + \Delta} \right\rangle_l$
(see Eq.~(\ref{eq4})). The results are shown in Fig.~\ref{fig6}.
For $n=2$ the numerical data are in a good agreement with the theory
for a harmonic chain. For $n=4,8$ our numerical data show the dependencies
rather similar to the case of a harmonic chain with slight vertical shift 
which can be attributed to the modified values of $\bar{\omega}$
discussed above. We note that for the case of 
$T > 0$ the theoretical formulas (\ref{eq3}), (\ref{eq4}) are computed with
the expression (\ref{eq5}) for bosons at finite temperature.
Since the value of $\tau$ in Fig.~\ref{fig6} is  rather large
there is no significant difference between the theoretical expressions for
$T=0$ and $T=\hbar/\tau$.

\begin{figure}[htbp]
\begin{center}
  \includegraphics[width=.52\linewidth]{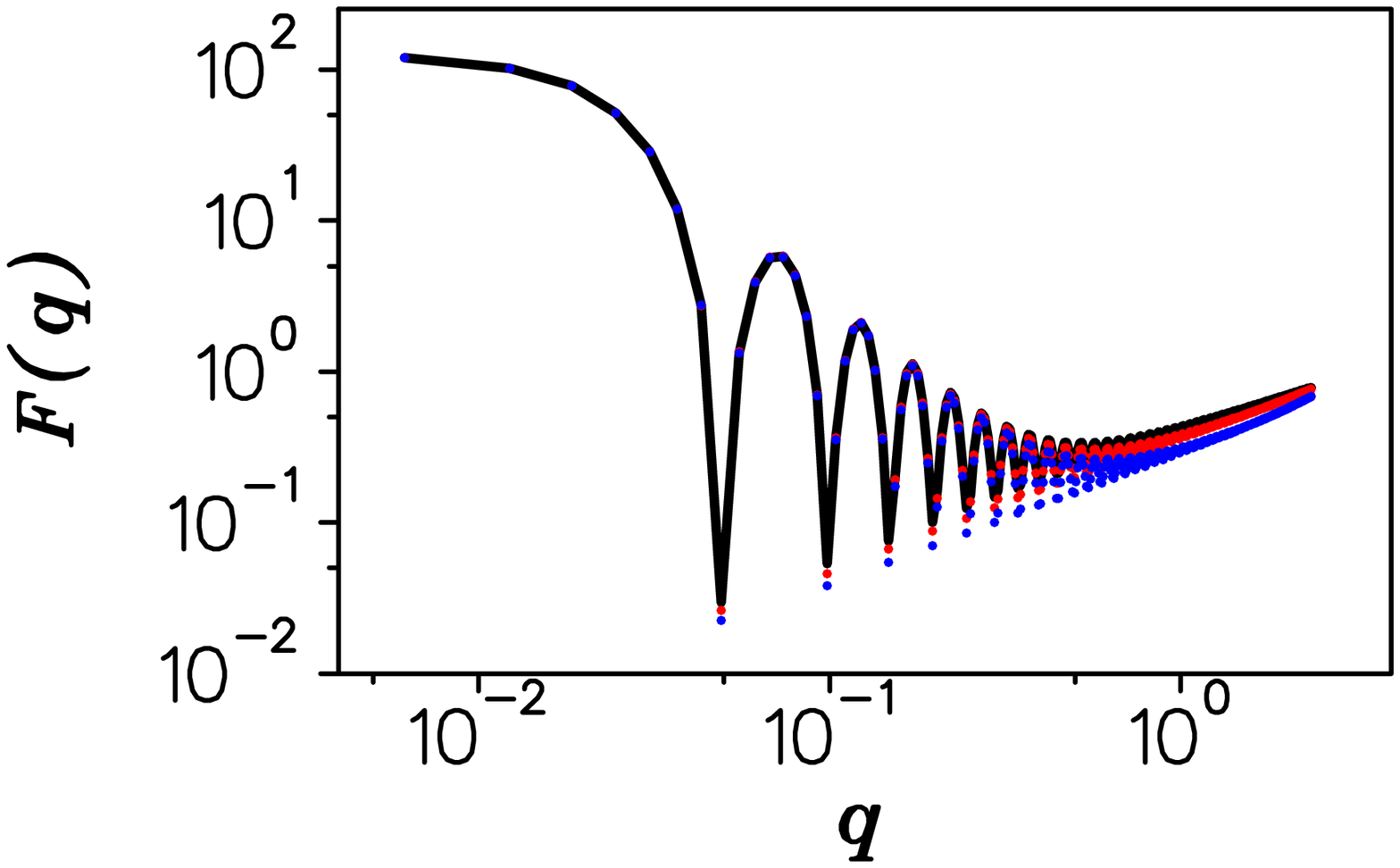}
  \includegraphics[width=.43\linewidth]{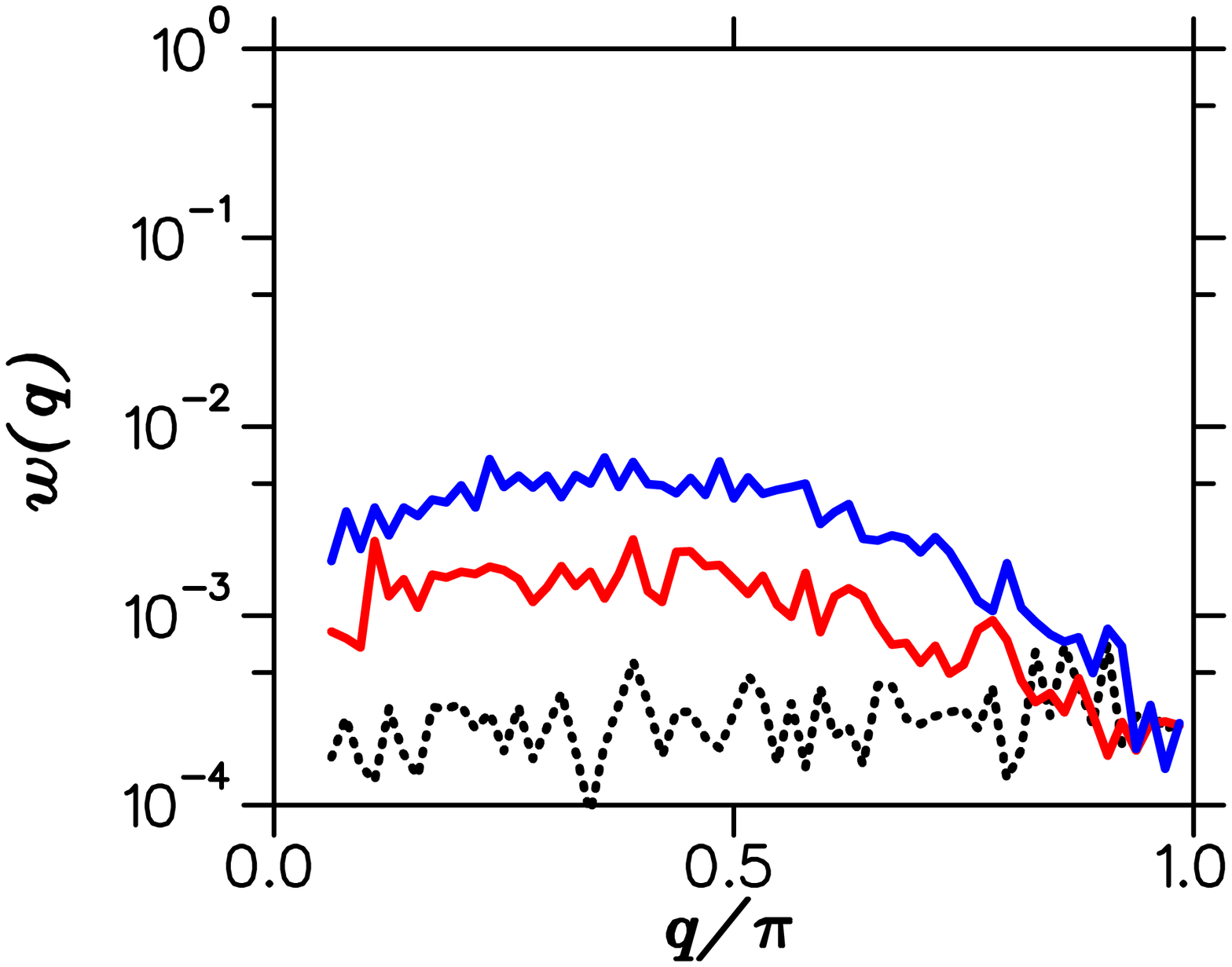}
\end{center}
\vglue -0.3cm
  \caption{(Color online) Left panel: formfactor $F (q)$, see Eq.(\ref{eq6}), 
  for $n=2$ (black curve),
  $4$ (red/gray points), $8$ (blue/black points) (data practically coincide).
  Other parameters are as in Fig.~\ref{fig6} including $a=1$. 
  Right panel: admixture to normal modes from other harmonics. Red/gray and blue/black
  curves show
  data for quartic and octic chains, the dotted black curve corresponds to linear
  chain and gives an estimate for a noise level. Data correspond to $\hbar=1$,
  $N = 64$, $\tau = 50$, $N_{\tau} = 500$, number of independent
  quantum paths is $10^4$. }
\label{fig7}
\end{figure}

An additional attempt to see deviations from a harmonic chain behavior is
performed by computing the form factor $F(q)$ of the chain given by Eq.~(\ref{eq6}).
However, the results presented in the left panel of Fig.~\ref{fig7}
show that all three chains with $n=2, 4, 8$ give very similar curves for $F(q)$
which practically overlap. 

The confirmation of similarity between these three types of chains
in the ground state is given by the direct computation of the 
correlator 
$ \left. K (q_1, q_2) \equiv \left\langle S_q^{} S_{q'} \right\rangle / 
 ( \left\langle S_q^{} S_q \right\rangle \left\langle S_{q'}^{} S_{q'}
  \right\rangle \right)^{1 / 2}$ 
which should be proportional to $\delta_{q_1,q_2}$ according to (\ref{eq3}).
Indeed, the numerical data show a strong peak at $q_1=q_2$
with a residual noisy level of $K$ at other $q_1 \neq q_2$
without any structural dependence on $q_1, q_2$.
This residual level  can be characterized by the total weighted admixture
of other modes to a given mode $q$ via
\begin{equation}
\label{eq15}
    w (q) = \sum_{q' \neq q} | \left\langle S_q^{} S_{q'} \right\rangle |^2 / |
  \left\langle S_q^{} S_q \right\rangle |^2 \; .
\end{equation}
This characteristic is shown in the right panel of Fig.~\ref{fig7} for $n=2,4,8$.
The admixture $w(q)$ increases with $n$ but still it remains rather small 
for strongly nonlinear lattices with $n=4,8$. This gives one more confirmation
that the quantum compacton vacuum is rather close to a harmonic one.

\begin{figure}[htbp]
\begin{center}
  \includegraphics[width=.43\linewidth]{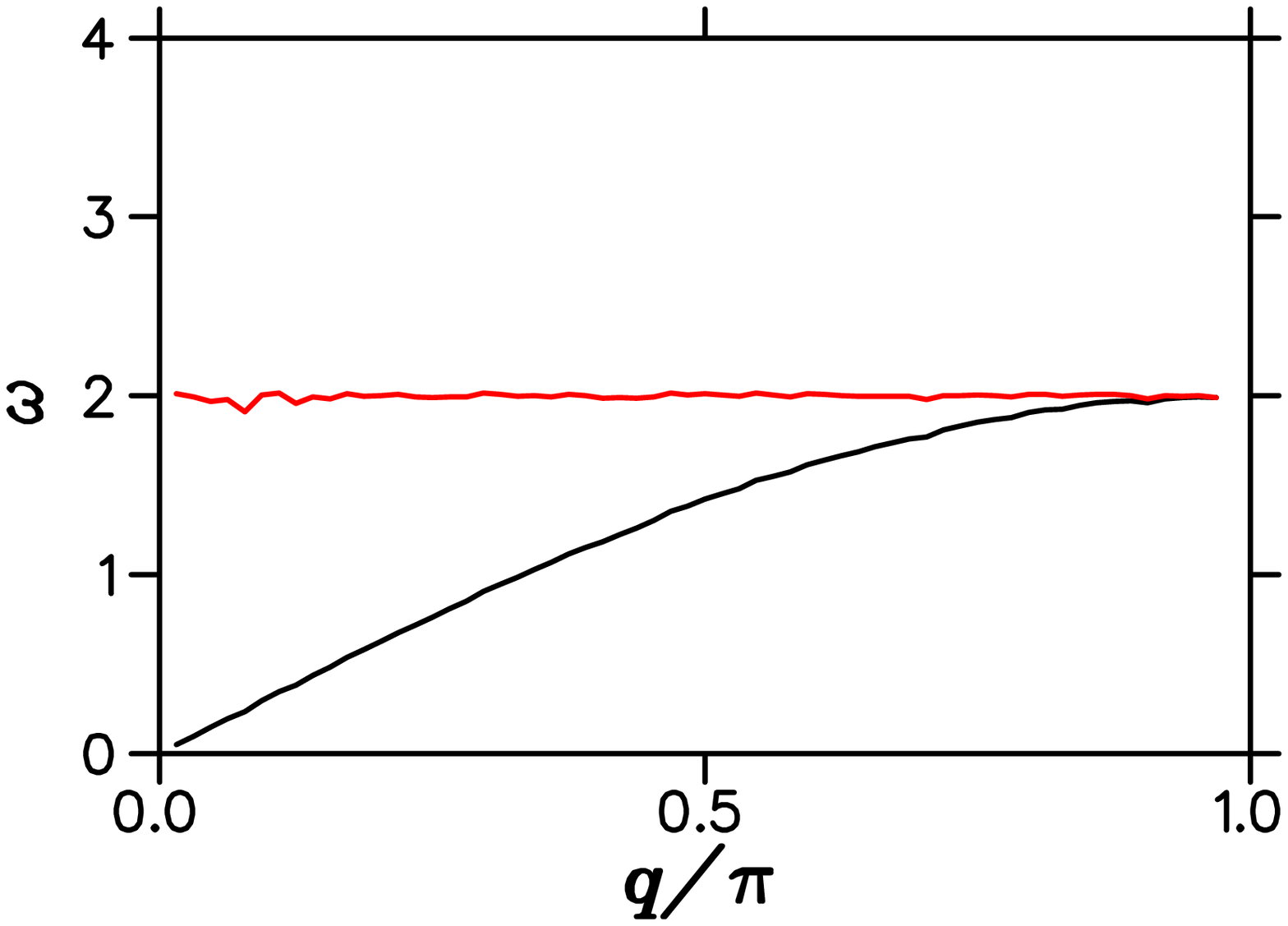}
  \includegraphics[width=.48\linewidth]{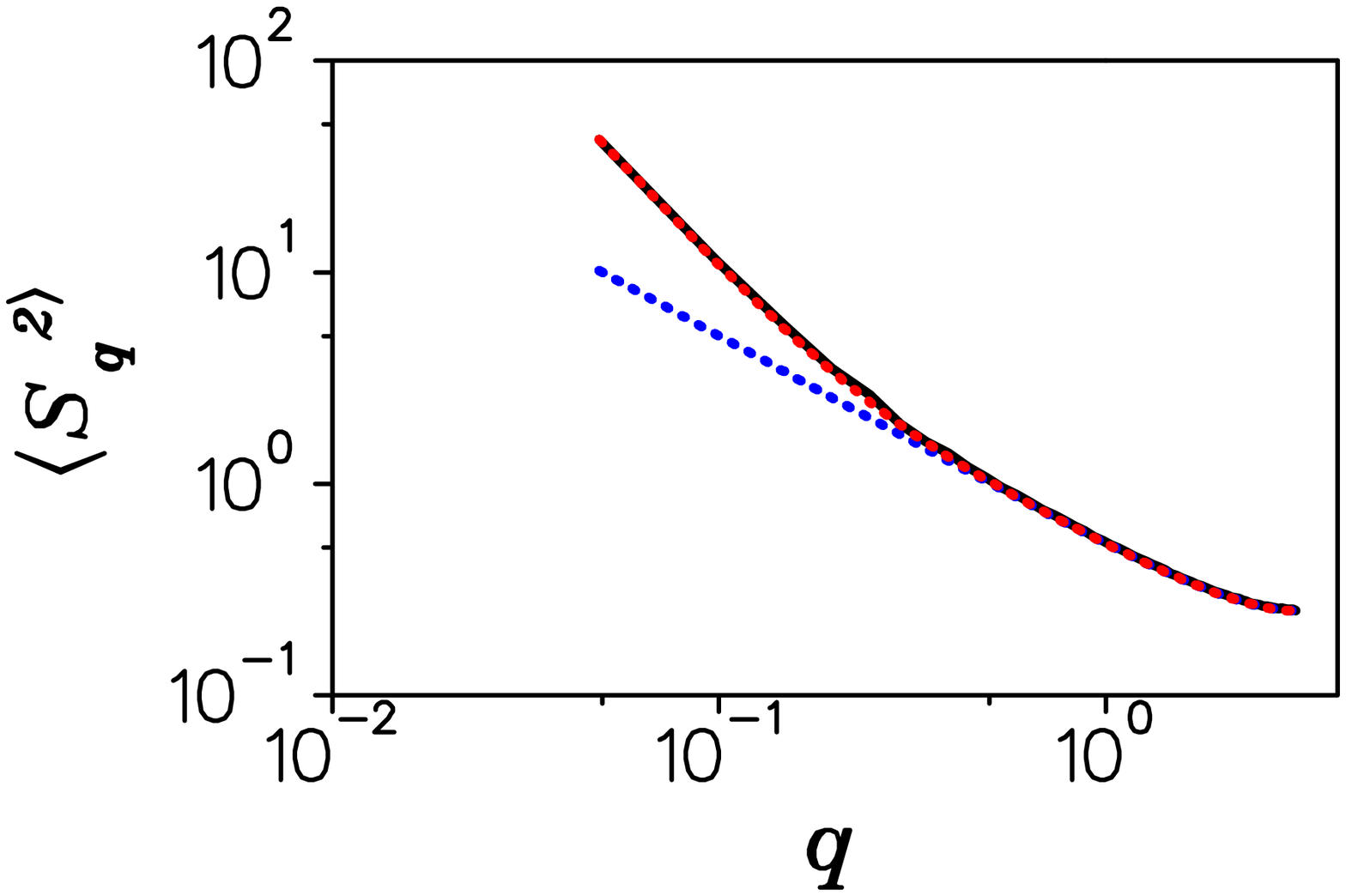}\\
  \includegraphics[width=.43\linewidth]{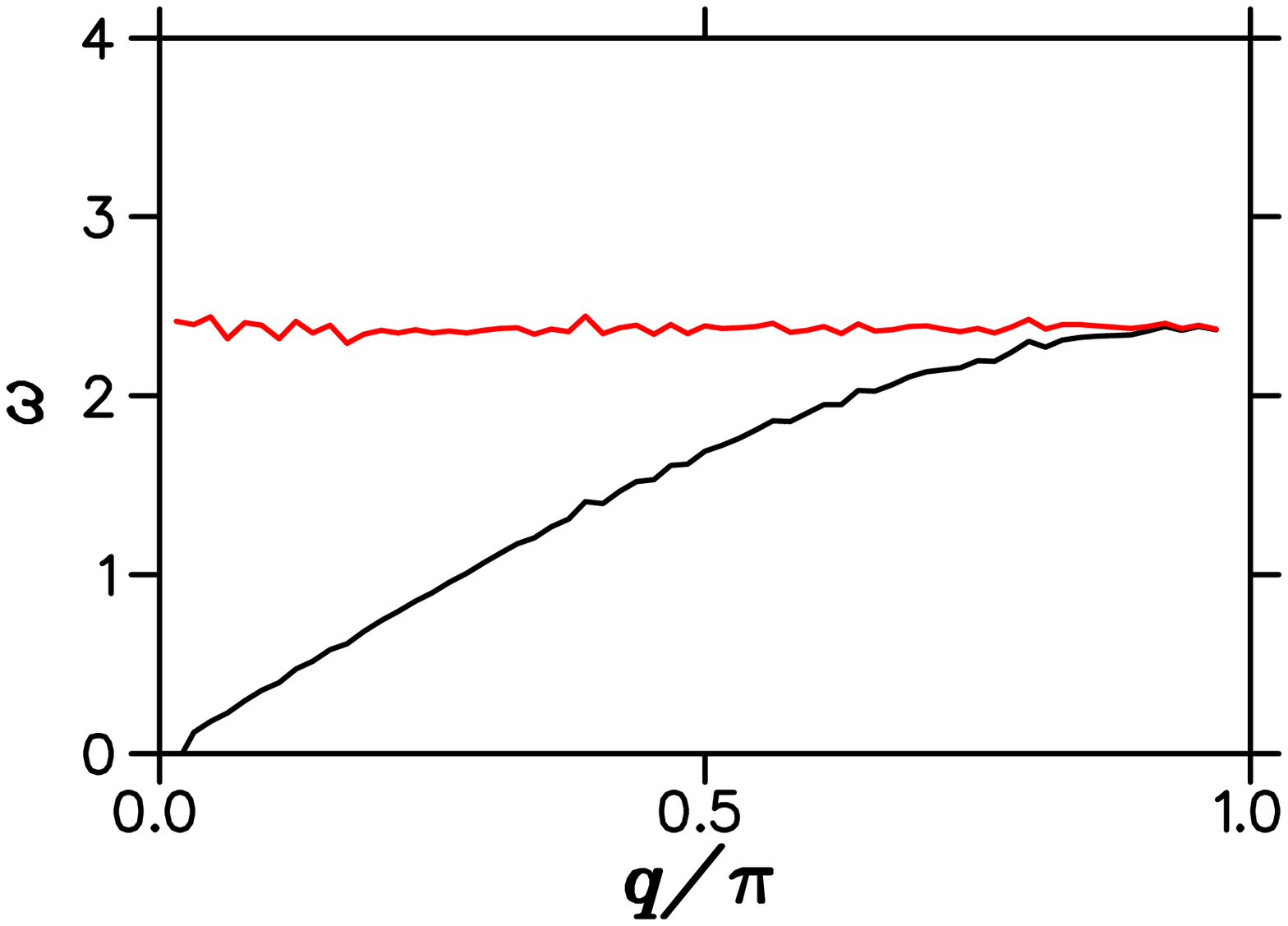}
  \includegraphics[width=.48\linewidth]{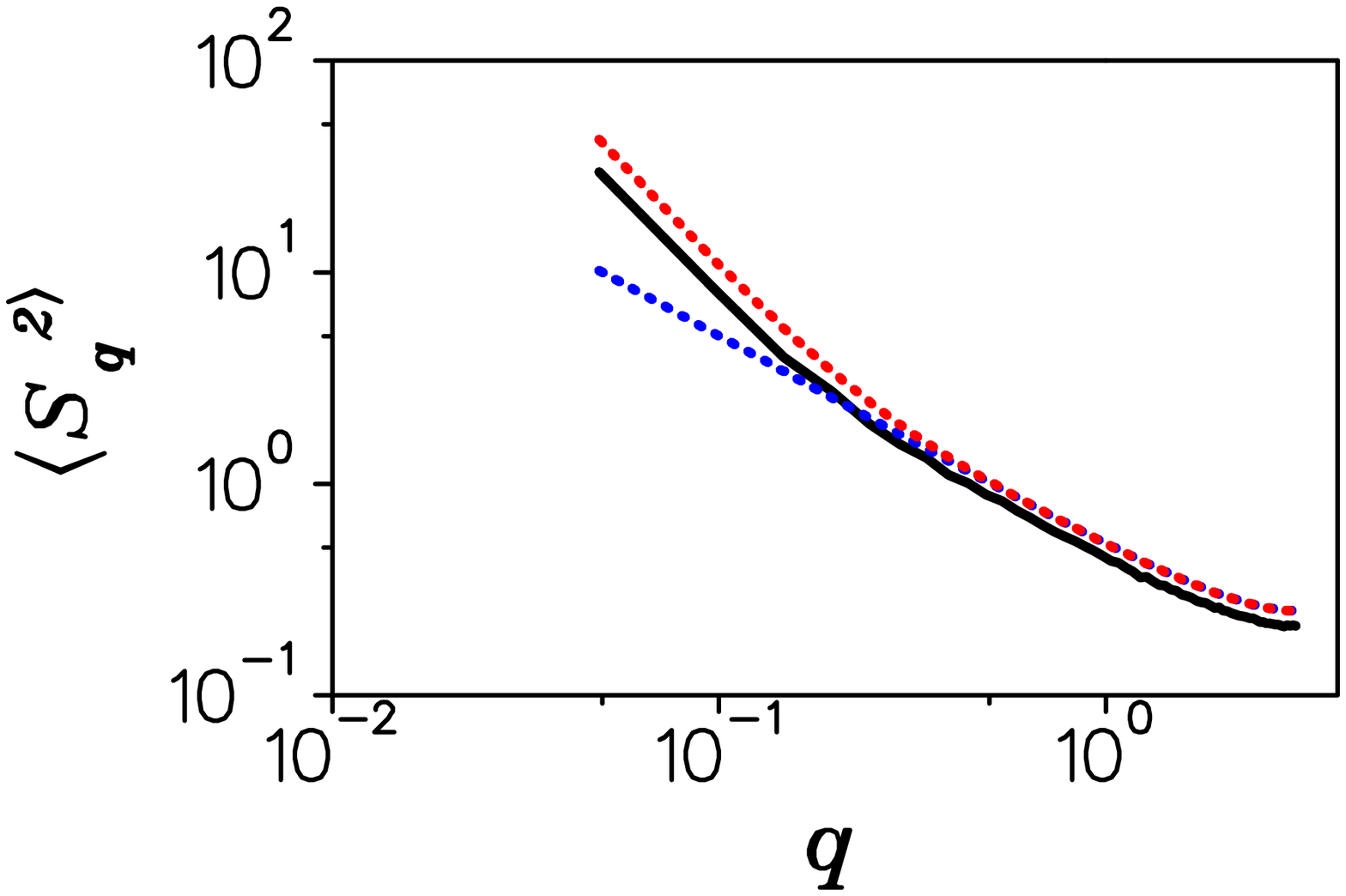}
\end{center}
\vglue -0.3cm
  \caption{(Color online) Left column: spectra of quantum phonon modes at $T=0.1$ for $n=2$
    (top) and $4$ (bottom) with the curves as in Fig.~\ref{fig3}.
  Right column: Amplitudes of phonon modes, see Eq.(\ref{eq3}) in the same
  order; red/gray and blue/black dotted
  lines give theoretical expectations for harmonic chain with temperature 
  $T = \hbar / \tau$ and $T = 0$, respectively.  Other parameters are
  $\hbar = 1$, $N_{\tau} = 200$, $\tau_{\max} = 10$. 
  Simulations are done with combined MA and MCD method with $10^4$ Metropolis 
  updates.
}
\label{fig8}
\end{figure}

At finite temperatures the spectrum of excitations $\omega(q)$
and the amplitudes of phonon modes $S^2_s$ are still well approximated by the 
theoretical dependence for a harmonic  chain as it is shown in Fig.~\ref{fig8}
at $T=0.1$ which is about two times larger than the excitation energy
for a mode with minimal frequency 
$\hbar \omega_{min}  = \hbar \pi/N \approx 0.049$. The situation is found to be
qualitatively the same when the temperature is increased up to $T=1$
keeping fixed other parameters of Fig.~\ref{fig8} even if 
at $n=8$ the splitting between the theoretical curve and the numerical data
becomes more visible (due to a similarity of these data  with Fig.~\ref{fig8}
we do not show them here).

\section{VI. Summary}

In this work we investigated the properties of collective modes in nonlinear
compacton chains. For the classical compacton chains we find that
the chain has sound modes which decay rather slowly due to nonlinear wave interactions
with the rate $\gamma \propto q^{\beta}$ with $\beta \approx 5/3$.
This decay rate is in agreement with the generic result of decay of sound
waves in one-dimensional nonlinear chains \cite{lepri}.

On local scales the classical dynamics in such compacton chains is
strongly chaotic. One could expect that this chaos may lead to nontrivial
properties of quantized chains. However, our extensive numerical studies
of quantum compacton vacuum show that it is characterized by
low energy phonon excitations which are rather similar to those
of a harmonic chain. The main difference is that the sound velocity
of these phonon modes depends on an effective Planck constant
as it is described by Eq.~(\ref{eq8}).

In a certain sense quantum effects suppress the signatures of classical chaos
in the ground state. Such a phenomenon is known for quantum systems with a few
degrees of freedom. For example, a ground state of a Sinai billiard 
can be rather well approximated by a Hartree-Fock trial function with 
one maximum so that the signatures of quantum chaos appear only 
in the semiclassical regime for highly excited states \cite{haake}. 
This is more or less natural for systems with few degrees of freedom.
Our case has infinite number of degrees of freedom
but in spite of that the quantum compacton vacuum remains rather
similar to a vacuum of a harmonic chain in which
the oscillator frequency is dependent of an effective Planck constant.
It is possible that certain signatures of such quasi-integrability
at low energies find their manifestations in a slow chaotization
of excitations in a quantum Newton's cradle observed in experiments
\cite{dweiss}. However, we should note that such a statement can be considered
only on a qualitative level of rigor since the model (\ref{eq1}) at $n=4$ or
$8$ gives only an approximate description of ball interactions (see discussion
at \cite{ahnert}).

However, let us note that our methods are well adapted 
for analysis of low energy oscillatory quantum waves. It is possible that other
approaches should be used to detect quantum chock wave type excitations
with large displacements between two parts of the chain
(in principle the energy of such type compacton like excitations
is not very high and is independent of the lattice size). 
Other methods should be used to
analyze such type of excitations.

It is possible that a degeneracy of chaos in a vicinity of quantum compacton
vacuum is linked to the space homogeneity of the model (\ref{eq1}).
Indeed, the ground state of the quantum Frenkel-Kontorova model
has much more rich properties \cite{qfk} appearing due to a presence of
periodic potential.

One of us (OVZ) is supported by
the RAS joint scientific program ``Fundamental problems in nonlinear dynamics''
and  thanks UMR 5152 du CNRS, Toulouse for hospitality during this work,
another (DLS) thanks Univ. Potsdam for hospitality at the final stage of this work.

%%**********************************************************************

\end{document}